\pdfoutput=1 
\documentclass{JINST}
\usepackage{subfigure}
\usepackage{url}

\usepackage{lineno}

\title{ATCA-based ATLAS FTK input interface system}

\author{Y. Okumura$^a$\thanks{Corresponding author.}~,
T. Liu$^b$, J.Olsen$^b$,
T. Iizawa$^c$, T. Mitani$^c$, T. Korikawa$^c$, K. Yorita$^c$,
A. Annovi$^d$, M. Beretta$^d$, M. Gatta$^d$,
C-L. Sotiropoulou$^e$, S. Gkaitatzis$^e$, K. Kordas$^e$, N. Kimura$^e$,
M. Cremonesi$^a$, H. Yin$^b$, and Z. Xu$^f$ on behalf of the ATLAS Collaboration\\
\llap{$^a$}University of Chicago Enrico Fermi Institute, \\
  5640 S Ellis Ave Chicago IL, 60637, USA\\
\llap{$^b$}Fermi National Accelerator Laboratory~\thanks{Operated by Fermi Research Alliance, LLC under Contract No. De-AC02-07CH11359 with the United States Department of Energy},\\
  P.O. Box 500, Batavia IL, 60510, USA\\
\llap{$^c$}Waseda University,\\
  3-4-1 Okubo, Shinjuku-ku, Tokyo, 169-8555, Japan\\
\llap{$^d$}INFN Frascati,\\
  Via E. Fermi, 40 00044 Frascati (Roma), Italy\\
\llap{$^e$}Aristotle University of Thessaloniki,\\
  Thessaloniki 54124, Greece\\
\llap{$^f$}Peking University,\\
  5 Yiheyuan Rd, Haidian, Beijing, China\\
E-mail: \email{yasuyuki.okumura@cern.ch}}

\abstract{The first stage of the ATLAS Fast TracKer (FTK) 
is an ATCA-based input interface system, where hits 
from the entire silicon tracker are clustered and 
organized into overlapping $\eta$-$\phi$ trigger towers 
before being sent to the tracking engines. 
First, FTK Input Mezzanine cards receive hit data 
and perform clustering to reduce data volume. 
Then, the ATCA-based Data Formatter system will 
organize the trigger tower data, sharing data among boards 
over full mesh backplanes and optic fibers. The board and system level 
design concepts and implementation details, 
as well as the operation experiences 
from the FTK full-chain testing, will be presented.}


\begin{document}

\section{Introduction} \label{sec:intro}
The ATLAS Fast TracKer (FTK) is a Phase-I upgrade program
for the ATLAS trigger system~\cite{cite::atlas_jinst, cite::ftk_tdr}. 
The ATLAS baseline trigger system consists of hardware-based Level-1 trigger and 
software-based Higher Level Trigger (HLT). 
The FTK is a hardware-based tracking processor for HLT.
It performs global track reconstruction after each Level-1 trigger accept signal,
in order to enable the HLT algorithms to have early access to tracking information.
FTK will use the entire silicon tracking system of about 100 million channels, 
consisting of the silicon pixel (Pixel) and 
micro-strip detectors (semiconductor tracker; SCT),
as well as the new Insertable B-Layer (IBL) pixel detector~\cite{cite::ibl_tdr}.
The availability of the tracking for entire detector, 
as well as the vertex information at the beginning of HLT 
will aid in the improvement of HLT algorithms, such as b-tagging and tau-identification.

\begin{figure}[tbp]
\centering
\includegraphics[width=.55\textwidth]{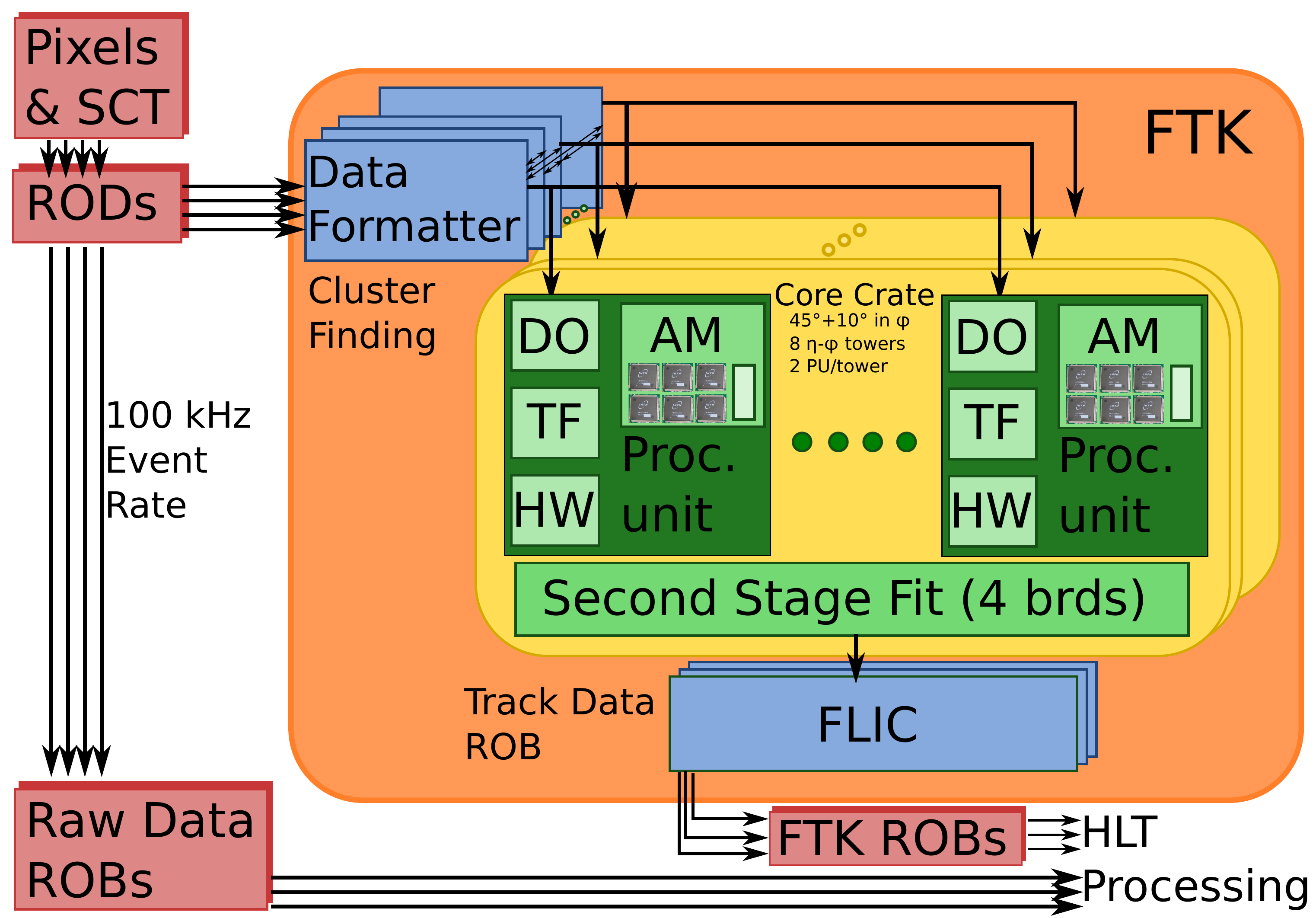}
\caption{
Functional sketch of FTK system. 
FTK is a pipeline of dedicated electronics, consisting of three main parts:
(1) Data Formatter for clustering and data formatting prior to the tracking,
(2) Core crate for pattern recognition and track fitter hardware units
(3) HLT interface crate (FLIC).}
\label{fig::ftk_system}
\end{figure}

\begin{figure}[tbp]
\centering
\subfigure[]{\includegraphics[width=.63\textwidth]{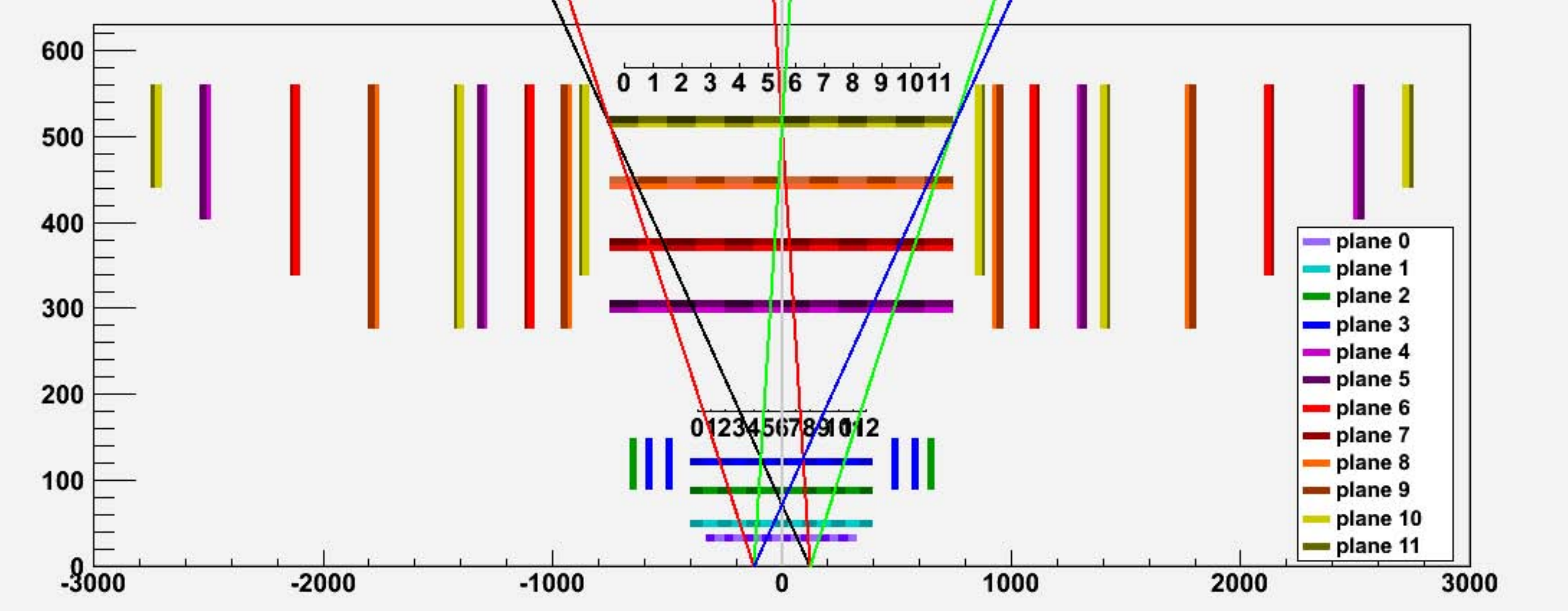}}
\subfigure[]{\includegraphics[width=.3\textwidth]{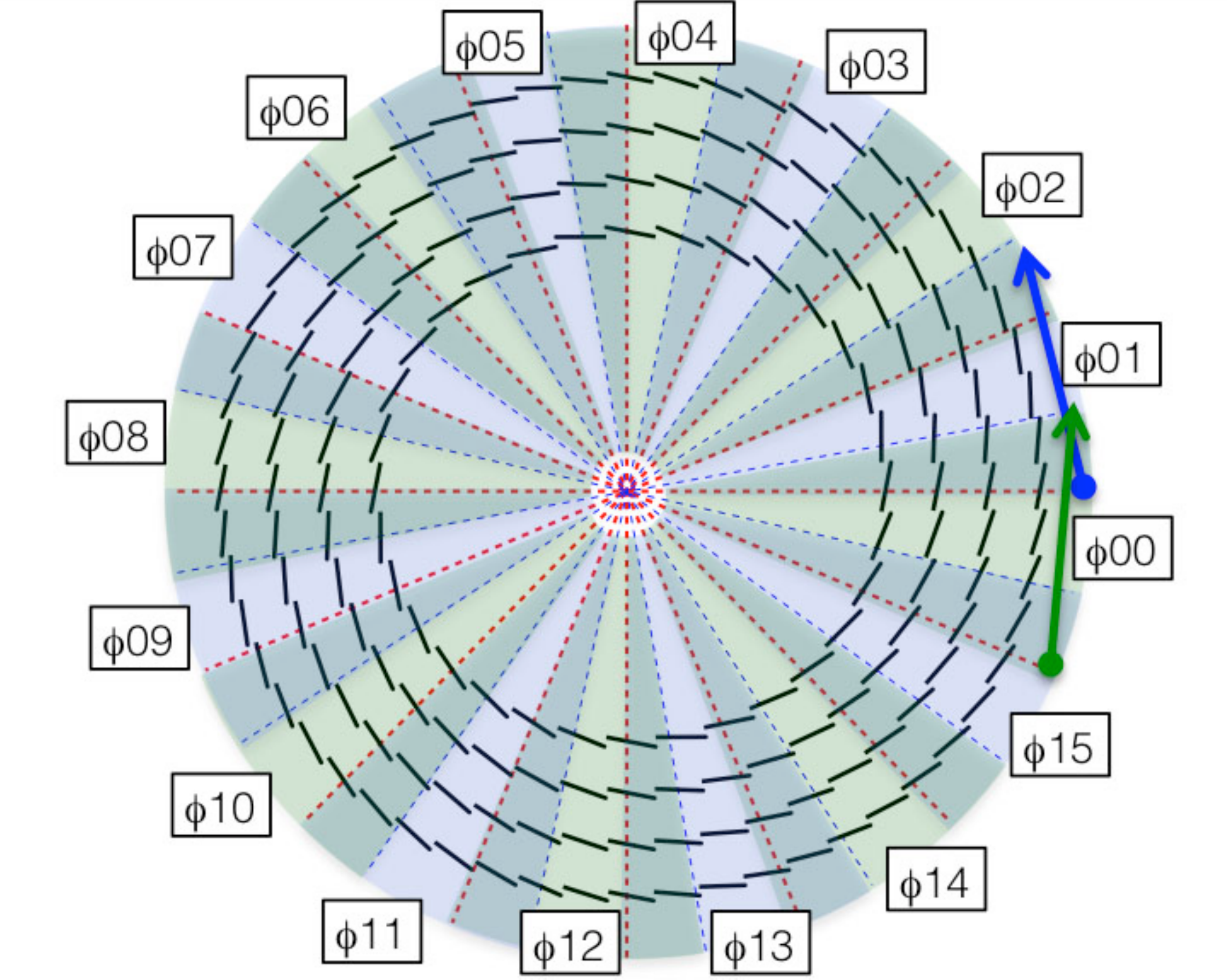}}
\caption{FTK 64 overlapping $\eta$-$\phi$ tower structure. 
The detector volume is divided into four sub-regions in the $\eta$ coordinate (a), 
and divided into 16 sub-sectors in the $\phi$ coordinate (b). 
}
\label{fig::eta_phi_tower}
\end{figure}

The FTK is a pipeline of dedicated electronics as shown in Figure~\ref{fig::ftk_system}.
FTK is highly parallel to deal with the large input data rate, as well as the hit combinatorics 
due to high occupancy at higher luminosity. 
The system is segmented into 64 overlapping $\eta$-$\phi$ trigger towers (Figure~\ref{fig::eta_phi_tower}), 
with its own pattern recognition and track fitter hardware units. 
The towers must overlap to avoid inefficiency at tower boundaries
due to the finite size of the beam luminous region in the $z$ coordinate
and the finite curvature of charged particles in the magnetic field. 
Considering the parallel processing structure, 
the ATLAS FTK requires its first stage to perform first clustering and then data formatting while data are read.
The clustering reduces the data volume and improves the coordinate resolution.
Data formatting is then needed to distribute the clusters to 
appropriate trigger $\eta$-$\phi$ towers, mapping 
the silicon detector readout structure to the FTK $\eta$-$\phi$ structure.
The system receives raw hits from Pixel, SCT and IBL ReadOut Drivers (RODs)
after the Level-1 trigger decision,
and delivers the data to the tracking processor units after clustering and data formatting.
This system is denoted as an ATCA-based ATLAS FTK input interface system in this paper. 

The following sections will present
the system-level and board-level design concept (Section~\ref{sec:design}),
the implementation details for data formatting, clustering, and user control interface (Section~\ref{sec:implementation}), and status of the system-level demonstration
of the test system at CERN (Section~\ref{sec:demonstration}).

\section{Design concept} \label{sec:design}
\begin{figure}[tbp]
\centering
\includegraphics[width=.6\textwidth]{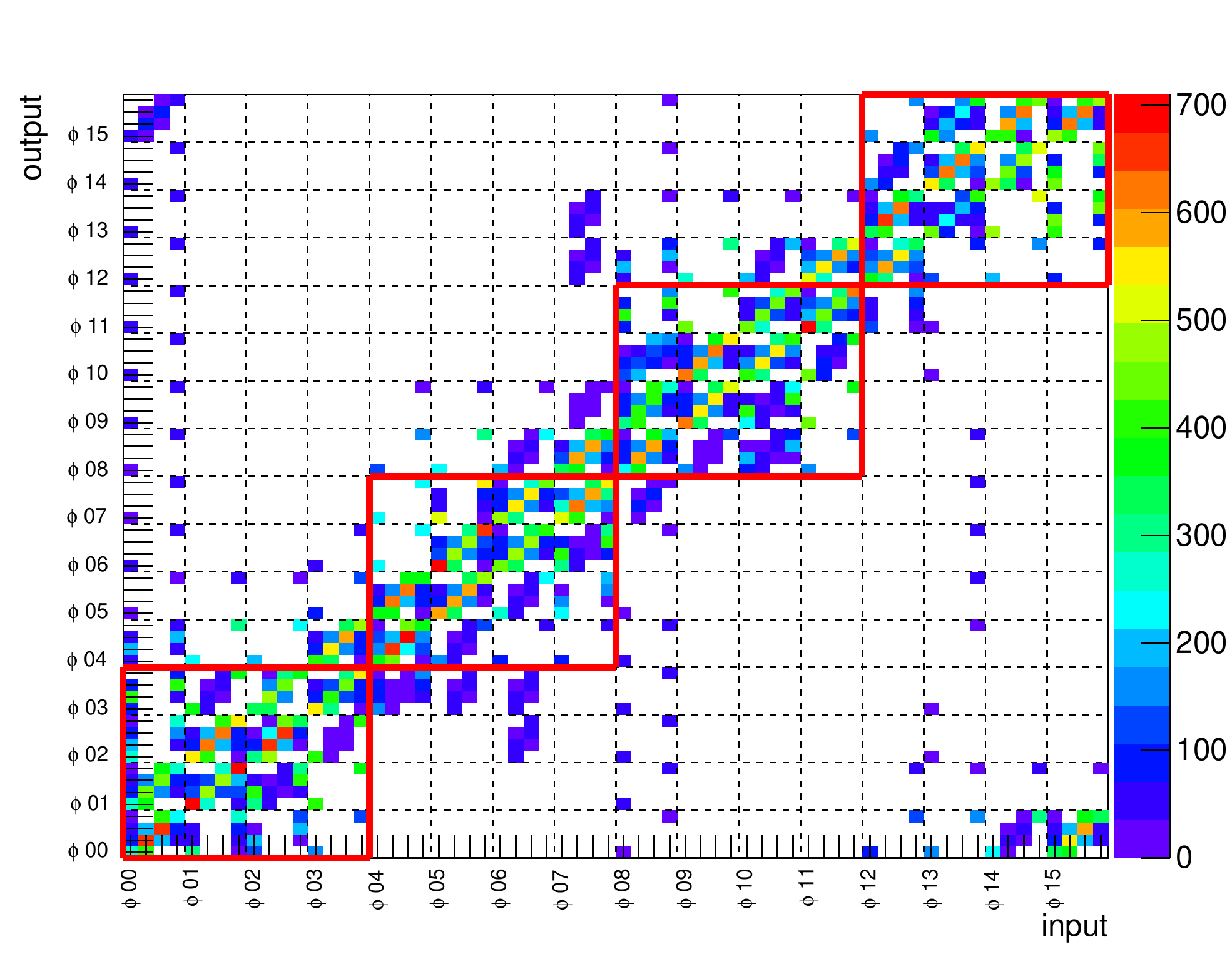}
\caption{
This $64\times64$ matrix shows the data sharing among the FTK $\eta$-$\phi$ towers 
in the data formatting operation.
The red boxes indicate the assignment of towers 
to 4 DF crates to minimize inter-crate data sharing.
The color scale indicates the number of clusters shared between towers per event with arbitrary scale.}
\label{fig:sharing_matrix}
\end{figure}

The system-level design of the input interface system is dominated by data formatting requirements.
We analyzed the required data sharing between the input and output of the system, using real beam data with the actual readout cable mapping.
The raw silicon hit data are input to the FTK system over about 
400 fibers from silicon RODs (input), and sent 
to the FTK $\eta$-$\phi$ towers through about 1,100 fibers (output).
The input fibers are assigned to output FTK $\eta$-$\phi$ towers,
so as to minimize the data sharing in the analysis.
Figure~\ref{fig:sharing_matrix} shows data sharing that is
needed among the FTK $\eta$-$\phi$ towers in the $64\times64$ matrix
after optimizing the input link assignments.
Off diagonal elements represent the data sharing between different towers.
The data sharing among trigger towers must be complex due to needs for
remapping the silicon detector readout structure 
into 64 FTK trigger $\eta$-$\phi$ tower partitioning and 
overlapping between neighboring towers. 

\begin{figure}[tbp]
\centering
\subfigure[]{\includegraphics[width=.25\textwidth]{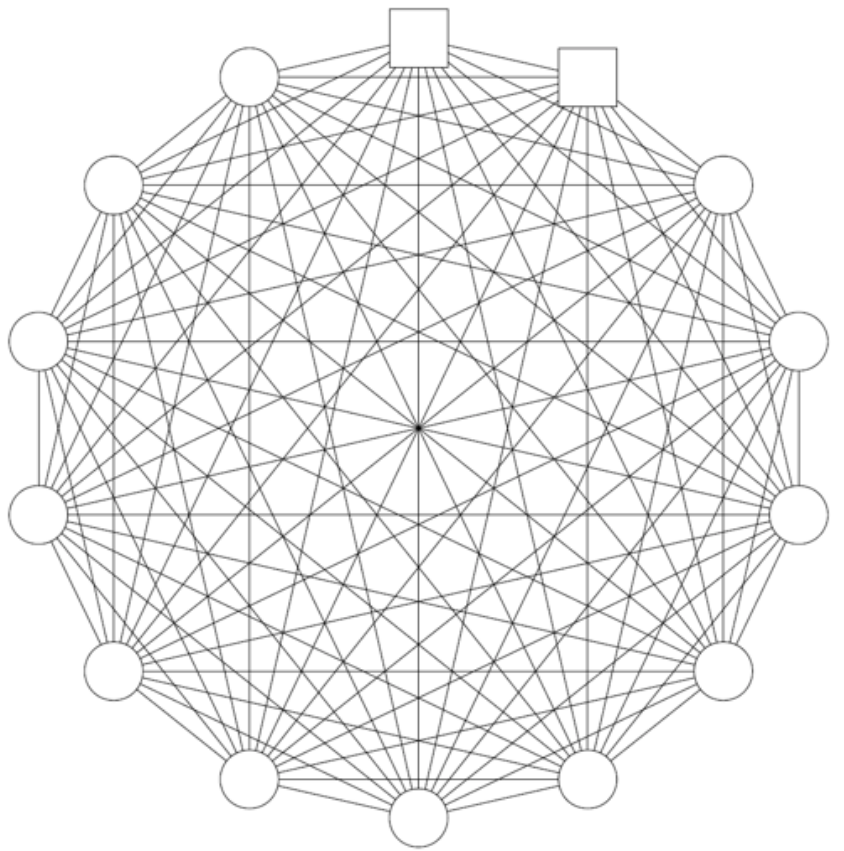}}
\subfigure[]{\includegraphics[width=.45\textwidth]{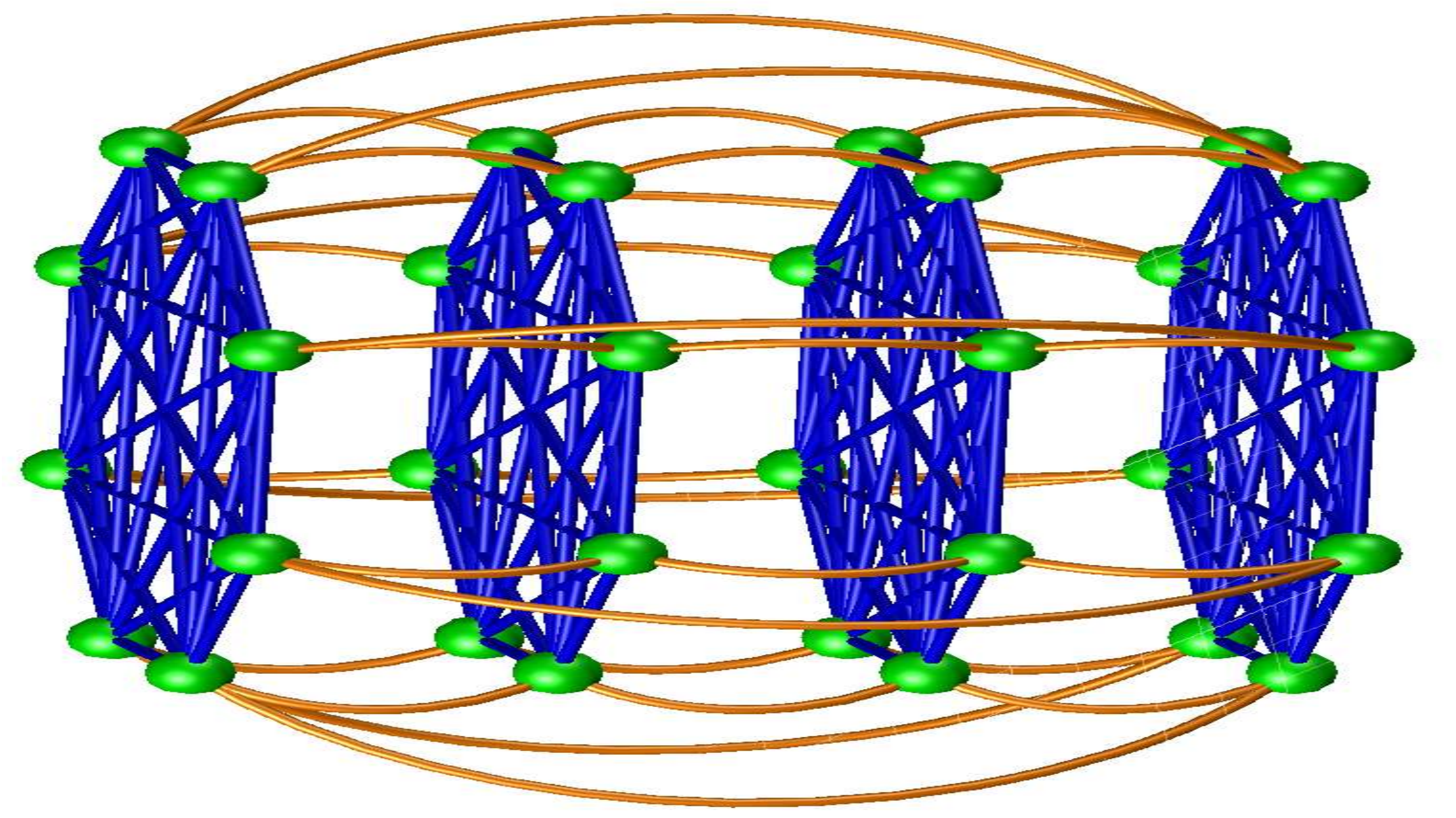}}
\caption{(a) The fabric interface connections in 14-slot full mesh ATCA backplane. 
Each line represents a multi-lane bidirectional channel rated for up to 40~Gb/s. 
(b) A graphical depiction of the 32 boards (in green) and high speed interconnect lines in four crate system. 
Blue lines represent backplane data paths.
Orange lines represent inter-crate fiber links.}
\label{fig:df_network_topology}
\end{figure}

The four red boxes in the matrix, which represent crate boundaries, 
show that a four-crate system fits the data formatting requirement.
Boards within each crate communicate over the backplane.
Fiber links need to be used when boards must communicate across crate boundaries. 
Further, our analysis shows that the data sharing between trigger towers
is highly dependent upon upstream cabling and detector geometry. 
The ideal data formatting hardware platform should be flexible enough 
to accommodate future expansion and allow for changes in input 
cabling and module assignments. 
Based on the requirements, 
the full mesh Advanced Telecommunication Computing Architecture (ATCA)~\cite{cite::atca}
backplane was found to be a natural fit. 
The Fabric Interface of the full mesh backplane enables high speed 
point-to-point communication between every slot (Figure~\ref{fig:df_network_topology} (a)), 
with no switching or blocking. Each line in this diagram represents a channel,
consisting of up to four bidirectional lanes, 
which runs at a combined maximum speed of up to 40~Gb/s. 
Assuming that we use 32 boards for the data formatting engines 
to deal with all 64 FTK $\eta$-$\phi$
towers, Figure~\ref{fig:df_network_topology} (b) shows a network topology 
that meets the data formatting requirements in the four full mesh backplanes.

\begin{figure}[tbp]
\centering
\includegraphics[width=.6\textwidth]{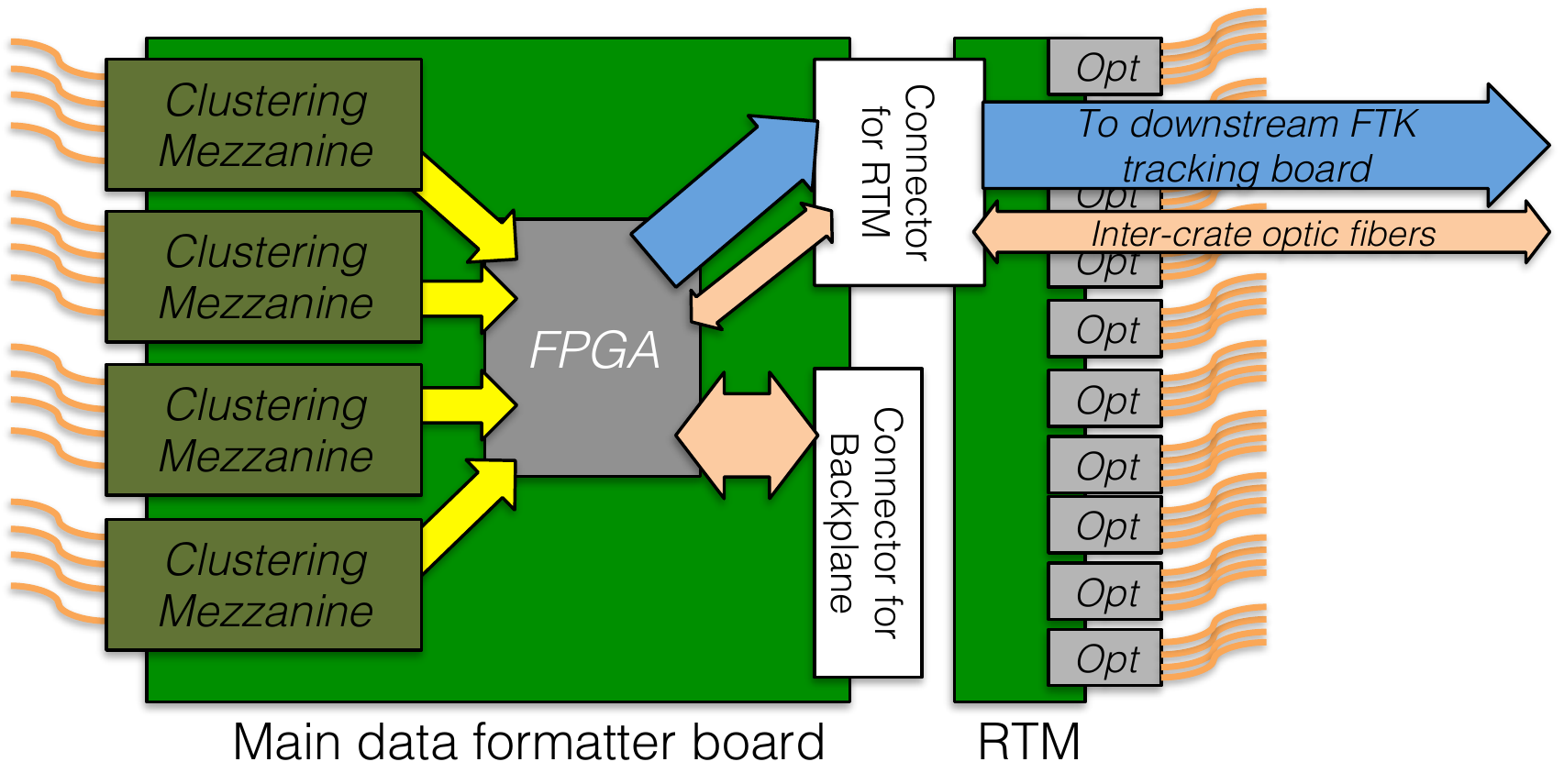}
\caption{Board-level design concept. Mezzanine card receives raw silicon hits from RODs 
and performs clustering prior to the data formatting to reduce the data volume.
The main board is used as a data formatting engine, interfaced to 
an ATCA full mesh backplane for point-to-point links, and pluggable transceivers on RTM for
inter-crate communication, as well as the links to downstream.}
\label{fig::board_level_design}
\end{figure}

Each board consists of three main components, as shown in 
the board-level design concept of Figure~\ref{fig::board_level_design}: 
1) four mezzanine cards (FTK Input Mezzanines; FTK IMs) to receive the input links and perform the clustering, 
2) a data formatting custom ATCA front board (Data Formatter front board; DF), and 
3) a rear transition module (RTM) to deal with the fibers for downstream links,
as well as the internal links beyond the crate boundary.
First the FTK IM receives the input raw silicon hit data.
Then clustering is performed by Field Programmable Gate Arrays (FPGAs) equipped on the mezzanine card.
The found clusters are then sent to the mother board FPGAs, 
which are used as the data formatting engines.
High speed serializer components in the FPGA are directly 
connected to the full mesh backplane fabric interface channels, 
and also to pluggable fiber transceivers located on the RTM,
used for interconnection among the DF shelves, and 
output links to downstream FTK processors.

\section{Hardware and firmware implementation} \label{sec:implementation}
32 DF boards will be used to handle 64 FTK $\eta$-$\phi$ towers,
where four FTK IMs connect to each DF board and 128 FTK IMs will be used in total.
We have developed the DF front board, RTM and 
FTK IM hardware, as well as the firmware for data formatting and clustering.
This section will describe details of the implementation for the DF in 
Section~\ref{sec:df_impl}, for the FTK IM in Section~\ref{sec:im_impl}, and 
for the user control interface in Section~\ref{sec:ui}.

\subsection{Data Formatter}\label{sec:df_impl}

\begin{figure}[tbp]
\centering
\subfigure[]{\includegraphics[width=.40\textwidth]{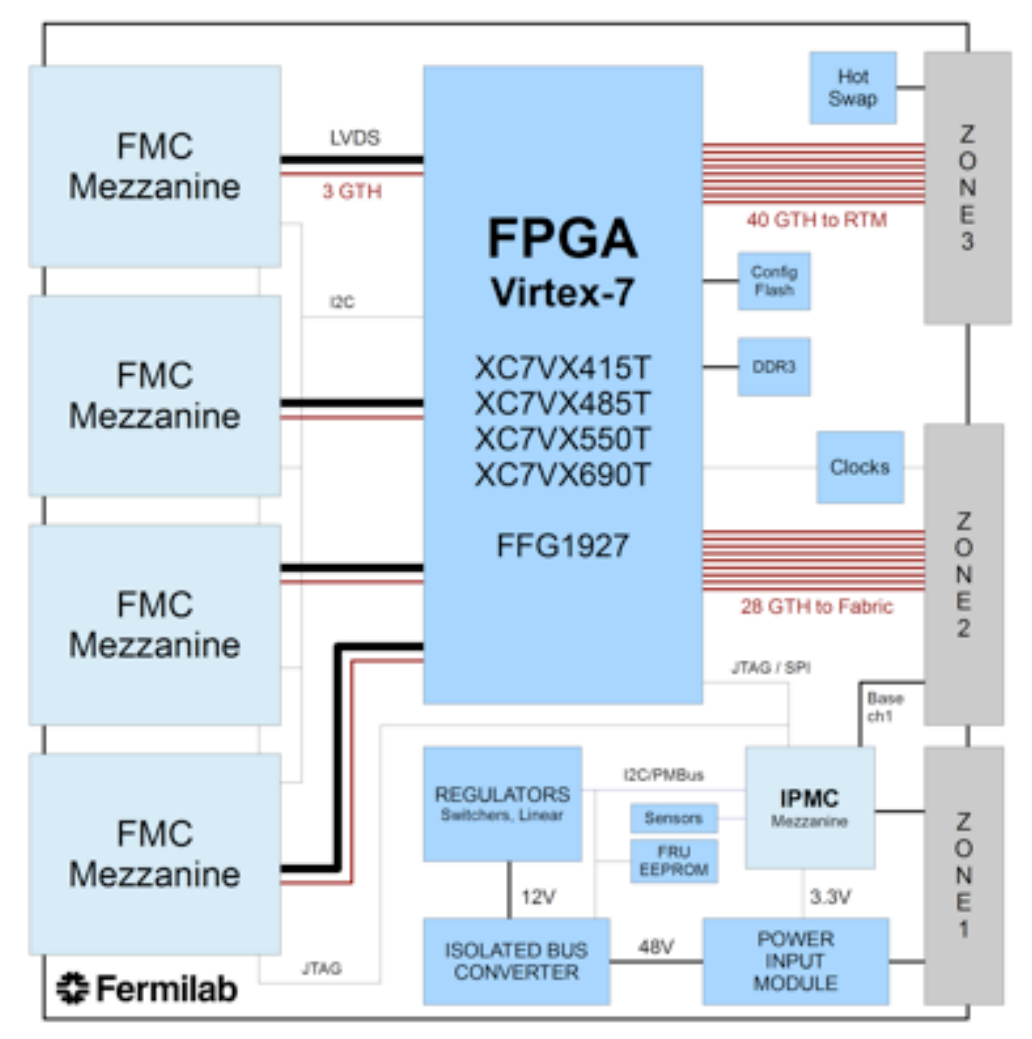}}
\subfigure[]{\includegraphics[width=.50\textwidth]{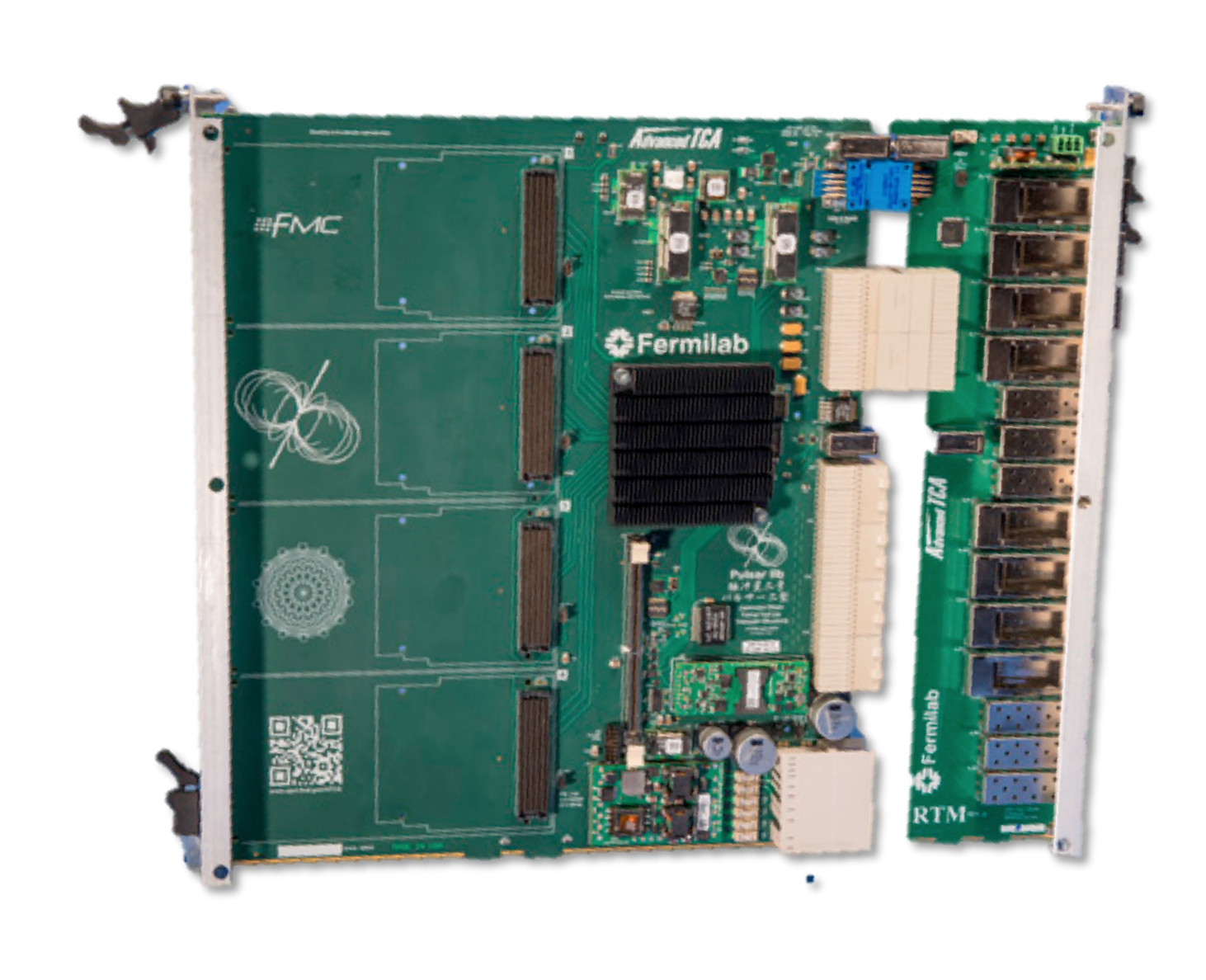}}
\caption{(a) Block diagram of the data formatter board. (b) Photo of the prototype board of DF board and the RTM.}
\label{fig:photo_and_block_diagram_data_formatter}
\end{figure}

The block diagram of the DF board is shown in Figure~\ref{fig:photo_and_block_diagram_data_formatter}.
The DF front board is designed around a Xilinx Virtex-7 FPGA (XC7V690T-2)~\cite{cite::Virtex7}.
Each board supports up to four FPGA Mezzanine Card (FMC)~\cite{cite::fmc} modules for the FTK IMs.
The FPGA has 80 high-speed built-in serial transceivers (GTH)~\cite{{cite::gth}}
which support data rates up to 10~Gb/s. 
Of these 80 GTH transceivers, 28 connect to the fabric interface, 
40 connect to the RTM, and the other 12 connect to the FMC connectors. 
The performance of GTH serial lines has been extensively tested and characterized 
with a bit error rate test (measured upper limit of bit error rate is less than $5\times10^{-16}$), 
as well as a receiver margin analysis (or a statistical eye scan).
The data traffic at 10~Gb/s has been proven to run stably on the DF board,
while 6~Gb/s is required for the FTK data formatting needs.

\begin{figure}[tbp]
\centering
\subfigure[]{\includegraphics[width=.25\textwidth]{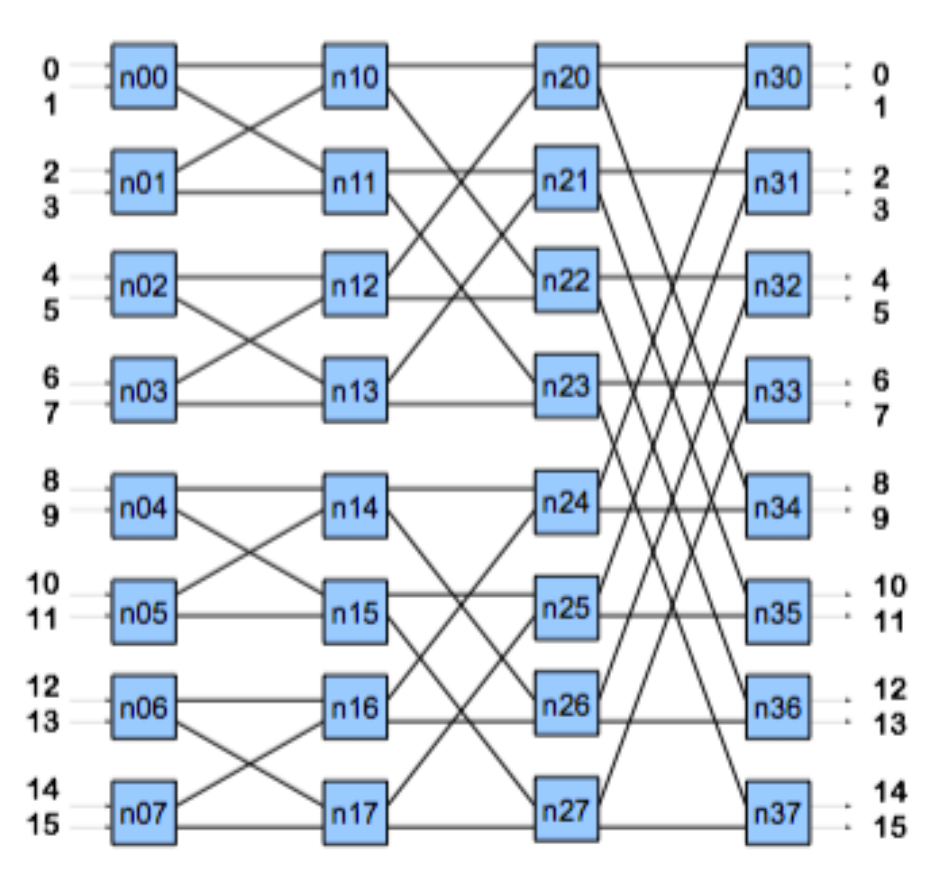}}
\subfigure[]{\includegraphics[width=.45\textwidth]{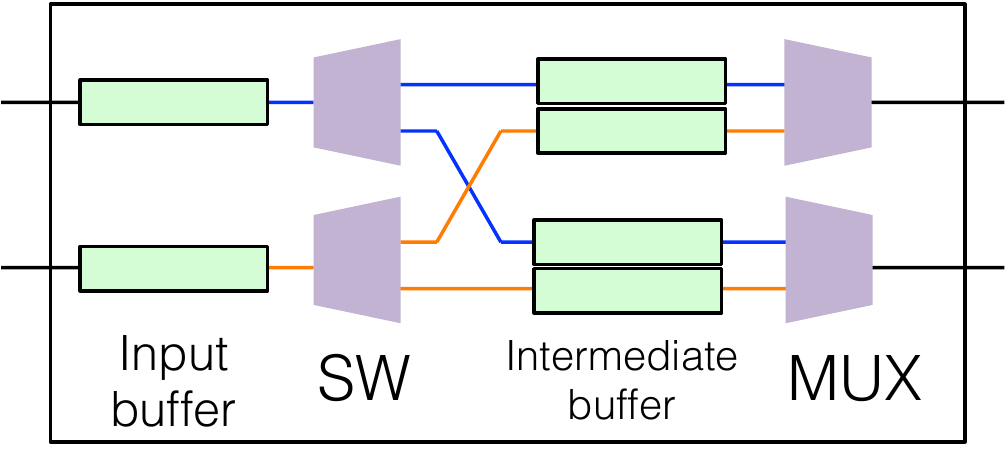}}
\caption{(a) An example of banyan network switch structure for 16-in-16-out implementation.
(b) Each 2-in-2-out switch element contains FIFOs for internal data buffering to minimize data blocking.}
\label{fig:data_formatter_implementation}
\end{figure}

The major challenge of the firmware implementation for the data formatting is 
to perform high bandwidth routing with low latency and flexible real time data sharing,
so that each board handles input lanes from the FTK IM (up to 16 lanes), output lanes to 
downstream FTK tracking boards (38 lanes), and internal communication lines connecting (18 lanes)
to the full mesh ATCA backplane and the inter-crate optic fibers.
A switch architecture with a banyan network is selected as the solution.
Figure~\ref{fig:data_formatter_implementation} (a) shows an example of the architecture 
with 16 input and 16 output lanes (16-in-16-out). 
Each switch element of the matrix structure contains first-in-first-out 
memories (FIFOs) for internal buffering  besides switching and multiplexing functions 
(Figure~\ref{fig:data_formatter_implementation} (b))
so that data blocking can be minimized with manageable resource usage.
The actual resource usage is approximately 70\%, where 
the firmware is equipped with three of 32-in-32-out switches.

The DF has the nickname Pulsar II. Although it was originally motivated by the 
FTK data formatting needs, it has proven to be generally useful in scalable system where
highly flexible non-blocking, high bandwidth board-to-board communication is required.
An example is the Level-1 silicon-based tracking trigger R\&D for CMS, where the full mesh
backplane is used effectively for sophisticated time-multiplexing 
data transfer schemes~\cite{cite::pulsariib_web,cite::pulsariib_design,cite::pulsariib_test}.

\subsection{FTK Input Mezzanine for clustering} \label{sec:im_impl}

\begin{figure}[tbp]
\centering
\subfigure[]{\includegraphics[width=.42\textwidth]{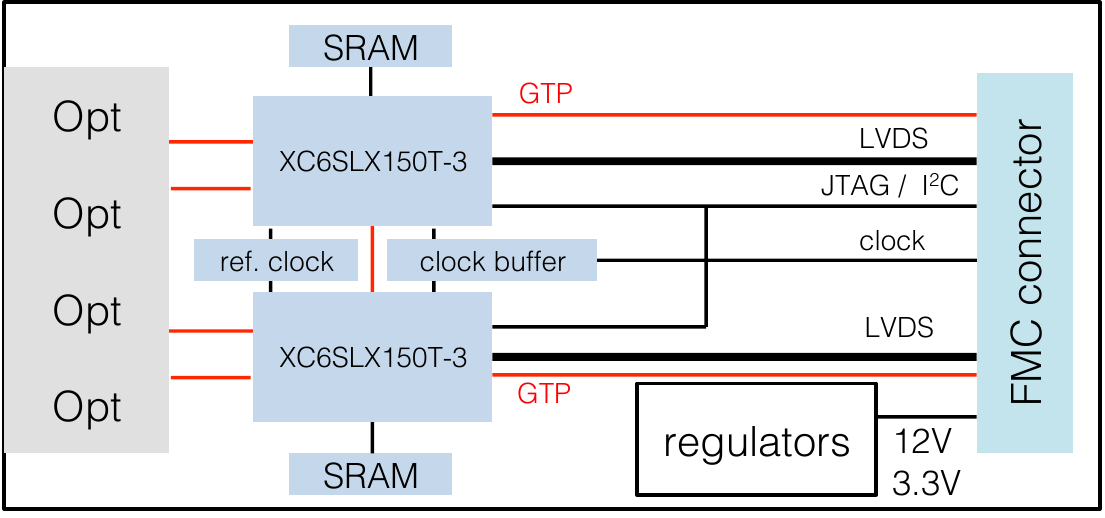}}
\subfigure[]{\includegraphics[width=.45\textwidth]{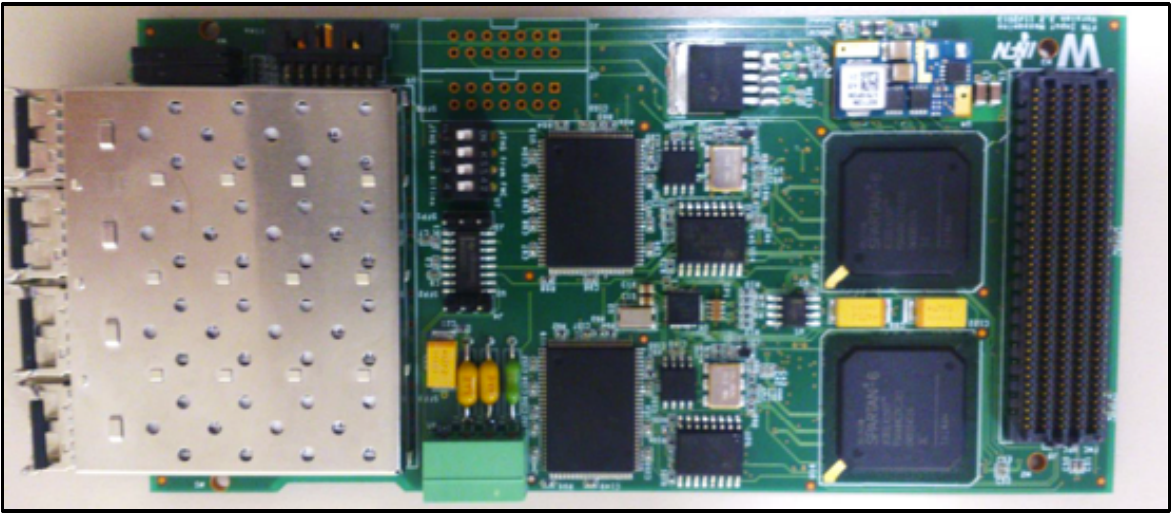}}
\caption{(a) Block diagram and (b) photo of the prototype board of the FTK IM.}
\label{fig:photo_and_block_diagram_ftkim}
\end{figure}
The FTK IM is a custom FMC module and a mezzanine card compatible with the DF mother board.
The block diagram of the FTK IM is shown in Figure~\ref{fig:photo_and_block_diagram_ftkim}.
It is equipped with two Xilinx Spartan-6 LX150T-3 FPGAs~\cite{cite::Spartan6},
four optical transceiver using the SFP to receive the input data from silicon RODs,
and an FMC connector to the DF board.
Each input SFP optical module connects to a GTP transceiver channel in the Spartan-6 FPGA 
running up to 3.2~Gb/s. S-LINK protocol~\cite{cite::slink} is used for the input links from RODs, 
running at 2.0~Gb/s. Each FPGA on the FTK IM connects to two optical modules. 
The FPGAs are used as the main engines for clustering, the logic of which is 
implemented with two decoupled stages: cluster finder and centroid calculators~\cite{cite::ftkclustering}.
Finally, the FTK IM sends the found clusters to the DF board 
by using 16 Double Data Rate (DDR) LVDS parallel buses running at 200~MHz over FMC connectors, 
allowing a data throughput of 32-bit words at 50~MHz with eight LVDS pairs per lane.
The interface is implemented according to the FMC standard, and provides 
a JTAG chain for FPGA programming, I$^2$C for control path, high-speed transceiver channels, 
power distribution, and clock distribution in addition to the LVDS parallel data bus.

\begin{figure}[tbp]
\centering
\includegraphics[width=.80\textwidth]{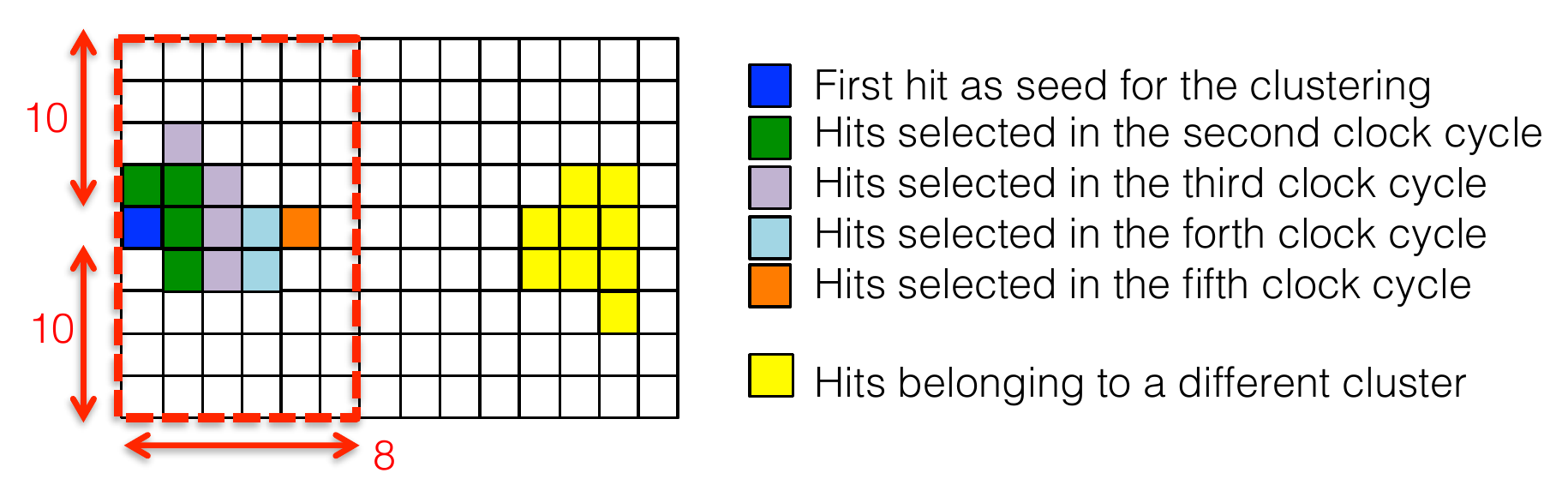}
\caption{Cluster finding procedure with the sliding window technique and two-dimensional hit selection method.}
\label{fig:clustering_implementation}
\end{figure}

It is a major challenge to maintain manageable processing time
of 2D cluster finding for the pixel sensors at the highest target luminosity of the 
FTK operation. 
In fact, the 2D structure of FPGA logic cells is suitable for mapping the 2D structure of sensors,
which allows the finder to avoid too many hit loops in the following way.
The finder logic starts with the first received hit (blue cell in Figure~\ref{fig:clustering_implementation}), 
the order of which has been sorted prior to the cluster finding.
It defines the cluster window ($21\times8$ pixel size) with respect to the 
first hit position so that the logic does not need to map entire pixel modules.
This is called the sliding window technique, and allows for an efficient resource usage.
Then the logic loads all hits within the window.
Once all hit data are loaded, 
the logic selects all the hits neighboring the first hits within one clock cycle.
This is the key step of the algorithm that avoids too many hit loops to maintain manageable processing time.
The hit selection is continued 
until there is no neighboring hit.
Eventually the selected hits are sent to the centroid calculator as a cluster. 
This way the full implementation fits in approximately 60\% of the available Spartan-6 resources.

\subsection{User control interface} \label{sec:ui}
Besides the main data formatting and clustering functions, 
user control interface has been implemented.
ATCA backplanes require two dedicated slots 
for hub switching boards (Hub Slots)
with two separate dual-star networks, called Base Interface for slow controls 
and Fabric Interface for high-speed transfers (up to 40G Ethernet).
A DF FPGA connects to one of the two Hub Slots
through Fabric interface with four lanes up to 40~Gb/s 
for the Ethernet communication. 
For the board control, we have adapted the CMS IPbus framework~\cite{cite::ipbus} to the system. 
UDP packets are transferred between external computers (IPbus clients) 
and the DF boards (IPbus devices) through a hub switching board.
Further I$^{2}$C bus signals for FTK IM may be controlled 
over the IPbus interface so that the control path from external computers is available
for both DF main boards and FTK IMs. 

\section{System-level demonstration} \label{sec:demonstration}
We have established an ATCA test-stand at CERN in order to demonstrate 
the system-level operation in terms of both the control path and data-flow.
Prototype DF boards and FTK IMs were integrated after the single board testing.
The control path with IPbus and I$^2$C has been established 
in the CERN General Purpose Network environment. 
One computer in the CERN network is setup as the IPbus client
on which an end-user application runs MicroHAL software. 
The IPbus transactions are routed via the Control Hub node,
which is packet-handling software that allows separation of hardware and control network,
for safe transport of IPbus commands.
For data-flow demonstration, first we tested communication functionality 
between neighboring boards with required speed: 
S-LINK (2.0~Gb/s) between RODs and FTK IM, 
LVDS parallel bus between FTK IM and DF (200~MHz DDR),
and S-LINK (6.0~Gb/s) between DF and downstream FTK tracking boards.
Then data-flow demonstration through the system
followed the interface functionality tests.
Either external optical transmitters which emulate RODs 
or test input patterns stored in FTK IM internal RAMs are used. 
The input event rate and data pattern can be adjusted, 
and the goal of the demonstration is to process simulation data with the highest target 
luminosity at 100~kHz input Level-1 trigger rate. 
The clustering algorithm and the data formatting functionality
are being tested in the test-stand.
As of the end of September, 2014,
we've demonstrated them with the simplest hardware setup:
one input at the maximum input speed of 100~kHz.
Scaling up the system is the next major milestone for the demonstration.

\section{Conclusion}
A clustering and data formatting solution for the ATLAS FTK has been developed, 
based on full mesh ATCA. We have developed a custom ATCA board
for data formatting engine (DF) and a custom FMC module for clustering (FTK IM).
Prototype boards and the firmware are being extensively tested to verify the design
and are being used for the system-level demonstration at CERN. 
The initial control path test and data-flow demonstration with the minimum setup has succeeded. 
The system-level demonstration with full FTK system functionality will be the next major milestone, 
for the FTK system commissioning planned in later 2015 during LHC Run2.

\end{document}